# SIMULATING ACQ/PEAK TARGET ACQUISITIONS WITH THE HST/FOS*

ROELAND P. VAN DER MAREL**
*Institute for Advanced Study,*
*Olden Lane, Princeton, NJ 08540, USA*

**Abstract.** A common target acquisition strategy with the Faint Object Spectrograph (FOS) on board the Hubble Space Telescope (HST) is to use a sequence of ACQ/PEAK stages. Standard sequences and exposure times are given in the FOS Instrument Handbook (Keyes et al. 1995). The success of a given strategy is difficult to predict a priori for targets other than point sources. This paper describes a FORTRAN program that simulates ACQ/PEAK acquisitions in a Monte-Carlo manner, for a variety of targets. The software is available from the FOS Instrument Scientists or from the author. It allows a proper quantitative assessment to be made of whether an acquisition will succeed, what exposure times should be used, and how the efficiency can be optimized.

## 1. Introduction

### 1.1 FOS ACQUISITION MODES

The positional uncertainty in the HST Guide Star Catalog is $\sim 0.5''$. Hence, some form of acquisition is required to properly position a target in the small apertures available on the FOS. Several acquisition modes are available for this purpose.

'Binary' acquisition (ACQ/BIN) is the most efficient mode. It acquires non-variable point sources with known magnitude and color in non-crowded fields. It is used mainly for stars. It yields a $1\sigma$ positional accuracy of $\sim 0.1''$.

The 'peak-up' acquisition mode (ACQ/PEAK) is more generally applicable. It can acquire any well-defined brightness maximum (e.g., a galactic nucleus) to essentially any positional accuracy. It can also be used to refine or verify the result of some other target acquisition mode, such as ACQ/BIN. It is time consuming, and can take up to 2.5 HST orbits.

In cases where the scientific target is faint or has a complicated geometry it is possible to acquire a suitable offset target within $\sim 30''$ (e.g., a nearby star), followed by a slew of the telescope to the scientific target. The offset must be known to high accuracy. In practice this requires that HST images be available and analyzed before the FOS observations are scheduled. Such images can be obtained in an 'early acquisition' (EARLY/ACQ) observation.

An 'interactive' acquisition mode (ACQ/INT) is also available. This mode uses real-time interaction with the telescope and requires the observer to be present a STScI. It is available only as last resort in exceptional cases. A final available acquisition mode

---

(ACQ/FIRMWARE) is used mainly for engeneering and calibration purposes, and is not recommended for general use.

## 1.2 ACQ/PEAK ACQUISITIONS

An ACQ/PEAK acquisition consists of a series of 'stages'. Each stage adopts a rectangular grid of SEARCH–SIZE–X by SEARCH–SIZE–Y points on the sky, with inter-point spacings of SCAN–STEP–X and SCAN–STEP–Y. An FOS aperture is positioned at each of the grid points, and the total number of counts is measured in some exposure time. The grid point with the most counts is adopted as new estimate of the target position (if requested, the algorithm can also adopt the position with the least counts). Subsequent stages in a sequence generally use smaller and smaller apertures, with each stage increasing the accuracy of the target positioning.

The observer who uses an ACQ/PEAK acquisition chooses the number of stages, as well as the aperture, the grid and the exposure time for each stage. The main constraint is that the final stage should achieve the positional accuracy required for the scientific objectives of the program. To optimize the efficiency, one wants to minimize the overhead and exposure times. The former are proportional to the total number of grid points in the sequence (30–40 sec per grid point), with an additional overhead per stage (a few minutes). The exposure times should be such that the probability of detecting the most counts at a grid point different from the one with the highest brightness (because of Poisson noise), is small. However, one doesn't want to expose longer than strictly necessary, especially for faint targets. Also important is that individual stages cannot be broken over different HST orbits. One thus wants to choose a sequence that fits efficiently into individual orbits.

The FOS Instrument Handbook lists 'standard' ACQ/PEAK sequences for acquiring a target in a given aperture with a given positional accuracy (see the example in Table I). The rule of thumb for choosing exposure times is that $\gtrsim 1000$ counts should be detected in the ACQ/PEAK aperture when placed at the position of the target. Stages in which the spacing of the grid is $\lesssim 0.15''$ are called 'critical'. For such stages the convolution with the HST PSF decreases significantly the brightness contrast between different grid points, and one should aim to detect $\gtrsim 10000$ counts. These rules are based on point source targets. Relations between FOS count rates and target magnitudes are given in the FOS Instrument Handbook.

The standard sequences and exposure times suffice for standard situations. In more complicated situations a variety of questions can arise, such as:

• What alternative sequences can be used if the standard sequences cannot be packed efficiently into HST orbits ?

• For extended targets (e.g., galactic nuclei) the brightness contrast between different grid points is always smaller than for point sources. Longer exposure times are thus required. Given a known extended target, how should the exposure times be chosen ?

• For faint targets (e.g., distant quasars) the standard exposure times can be very long, especially for critical stages. Can one use shorter exposure times, without increasing the risk of acquisition failure to an unacceptable level ?

• Does a given acquisition sequence work for a complicated target geometry (e.g., a point source on a background, or a point source in a crowded field) ?

• If the 0.1 aperture is used, or if high positional accuracy ($\lesssim 0.05''$) is required: How do the nature of the HST PSF and telescope jitter influence the acquisition ?

To answer the first question it is useful to have an extensive list of possible stages and sequences to choose from. For the rectangular and circular FOS apertures the geometrical constraints on the ACQ/PEAK parameters can easily be written in closed form. I have generated lists of ACQ/PEAK stages and sequences satisfying these constraints, applicable to various situations: (i) acquisitions consisting entirely of ACQ/PEAK stages; (ii) acquisitions in which one or two ACQ/PEAK stages are used to refine the result of an ACQ/BIN acquisition; and (iii) acquisitions in which an ACQ/PEAK stage is used to verify the result of an EARLY/ACQ, or of a re-using of ACQ/PEAK offsets obtained in a previous visit. The lists form an extension to the standard lists given in the FOS Instrument Handbook. They will be published elsewhere and can be obtained from the author upon request.

To address the remaining questions in the above list one needs software that simulates ACQ/PEAK acquisitions in a Monte-Carlo manner. The remainder of this document describes a FORTRAN program that I developed for this purpose, with some examples. The program is available upon request, either from the FOS Instrument Scientists or from the author. It is user-friendly and mostly self-explanatory.

## 2. Description of the ACQ/PEAK simulation software

### 2.1 OUTLINE

The simulation program works as follows. At each grid point in a stage the surface brightness of the target (see Section 2.1) is convolved with the HST/FOS PSF, and is integrated over the area of the aperture (see the Appendix). The total magnitude in the aperture is transformed to an expectation value for the number of detected counts, using the known HST/FOS efficiency. A random deviate is drawn from a Poisson distribution with this expectation value, to model the shot-noise. The program runs many Monte-Carlo simulations of the acquisition sequence. Diagnostics are returned about the success and accuracy of each simulated target acquisition. The initial telescope pointing offsets are drawn from a Gaussian distribution with a dispersion corresponding to the a priori uncertainty in the target coordinates.

### 2.2 THE SURFACE BRIGHTNESS DISTRIBUTION OF THE TARGET

Two choices for the surface brightness distribution $S(x,y)$ of the target are currently implemented in the software:

(1) A point source, possibly on a uniform background: $S(x,y) = A\delta(x,y) + B$, where $A$ and $B$ are constants and $\delta(x,y)$ is the delta-function.

TABLE I
Example: Four-stage acquisition into the 0.3 aperture.

| stage # | aperture | SEARCH-SIZE-X | SEARCH-SIZE-Y | SCAN-STEP-X (arcsec) | SCAN-STEP-Y (arcsec) | Crit. ? |
|---|---|---|---|---|---|---|
| 1 | 4.3 | 1 | 3 | — | 1.23 | N |
| 2 | 1.0 | 6 | 2 | 0.61 | 0.61 | N |
| 3 | 0.5 | 3 | 3 | 0.29 | 0.29 | N |
| 4 | 0.3 | 4 | 4 | 0.11 | 0.11 | Y |

Fig. 1. Schematic illustration of the aperture positionings on the sky for the four-stage ACQ/PEAK sequence listed in Table I. The relevant height of the 4.3 aperture in the first stage is the height of the diode array of the detector. The scale is different in each panel. At the beginning of each stage the target is known to be contained in a region with size equal to the aperture used in the previous stage (provided that the previous stage was successful), indicated by the thick curve. In an idealized situation the sequence yields a maximum positioning error of $0.08''$.

**(2)** The nucleus of an (elliptical) galaxy with the following parametrization for the major axis surface brightness profile:

$$S(x,y) = S_0 \, (r/r_b)^\gamma \left[1 + (r/r_b)^\alpha\right]^{(\beta-\gamma)/\alpha} \tag{1}$$

This provides a good fit to the central regions of galaxies observed with HST (e.g., Kormendy et al. 1994). The surface brightness behaves as $S \propto r^\gamma$ at small radii ($\gamma \leq 0$). It diverges for $\gamma < 0$ and has a core for $\gamma = 0$ (in which case $S_0$ is the central surface brightness). The surface brightness falls off as $S \propto r^\beta$ at large radii ($\beta < 0$). The break occurs at the scale radius $r_b$, with $\alpha$ measuring the sharpness of the break ($\alpha > 0$). The user can also specify the apparent flattening of the galaxy, and the angle between the major axis and the FOS $x$-axis (determined by the roll angle of the telescope).

Users who wish to simulate acquisitions of other or more general extended targets need only supply a FORTRAN subroutine which returns the surface brightness distribution $S(x,y)$ of the target.

## 3. Example

Consider the case in which observations are planned with the circular 0.3 aperture. Let a positional accuracy of $0.08''$ be sufficient for the scientific observations. For targets not-suited for a binary acquisition the four-stage ACQ/PEAK sequence listed in Table I is recommended in the FOS Instrument Handbook. Fig. 1 illustrates the aperture positionings on the sky for the stages in this sequence.

Fig. 2 shows the results of simulations of this acquisition sequence. The circle in each panel is the 0.3 aperture. The square is the region of $0.11''$ by $0.11''$ (i.e., SCAN–STEP–X by SCAN–STEP–Y as used for the final stage), within which the target is acquired in an idealized situation. The dots show for 500 Monte-Carlo simulations the target offset from the adopted telescope pointing position after the final stage.

The left panel shows the results for a point source target using the standard exposure times (see Section 1). The target falls inside or close to the ideal square for all simulations. The mean final positioning error is $0.047''$, the maximum is $0.079''$.

The middle and right panels show the results for a target that is more difficult to acquire: the nucleus of a (round) elliptical galaxy with a core within which the surface brightness levels off to a constant value. The parameters in eq. (1) were set to $r_b = 1''$, $\alpha = 2$, $\beta = -2$, $\gamma = 0$.

In the middle panel the exposure times were chosen as for the point source in the left panel, i.e., to give 1000 counts in the the first three stages and 10000 counts in the final critical stage. This yields poorer results than for the point source, which is more centrally peaked. The target falls outside the aperture in 6% of the simulations. The mean final positioning error is $0.072''$, the maximum is $0.184''$.

Better results are obtained when the exposure times are increased (which is feasible only when the galaxy is sufficiently bright). The right panel shows the results of simulations in which the exposure times are a factor 10 larger than in the middle panel. The mean final positioning error is $0.050''$, the maximum is $0.109''$. The target is now contained in the aperture for all 500 simulations.

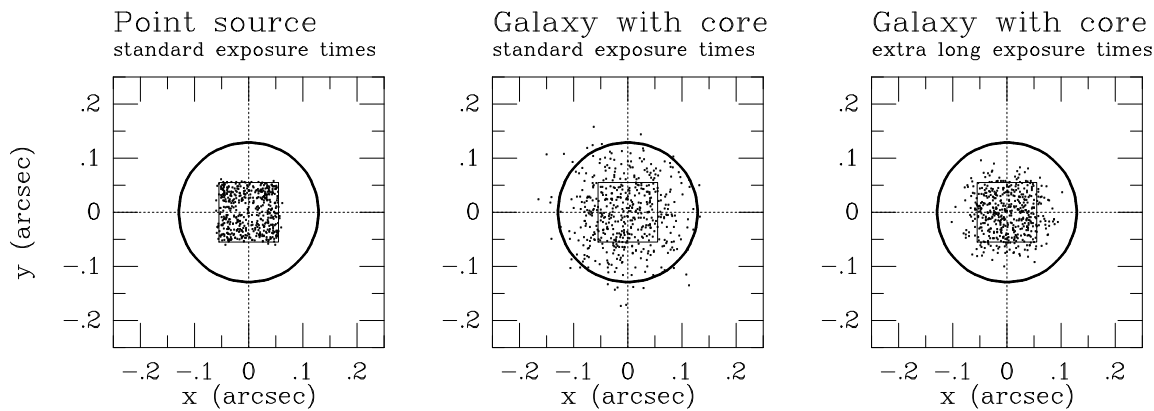

Fig. 2. Example results of the ACQ/PEAK simulation software as described in the text. Each panel shows the results of 500 Monte-Carlo simulations of the sequence plotted in Fig. 1. The circle is the 0.3 aperture. The square is the region within which the target is acquired in an idealized situation. In the left panel the target was chosen to be a point source. In the middle and right panels it was chosen to be the nucleus of an elliptical galaxy with a $1''$ core within which the surface brightness levels off to a constant value. The left and middle panels use the 'standard' exposure times, the right panel uses a 10 times longer exposure time. The results show that a proper acquisition requires longer exposure times for the galactic nucleus than for the point source.

Telescope jitter can cause small ($\lesssim 0.02''$) pointing errors at each step in a stage. Simulations for the sequence in Table I indicate that this does not lead to acquisition failure. There is a certain overlap between the apertures at different grid points in a stage (see Fig. 2). This reduces the probability that a grid point is adopted for which the aperture doesn't actually contain the target. Even if the wrong grid point is adopted, the target is usually recovered in the next stage, because it scans a region that is larger than the aperture size of the previous stage. Therefore, telescope jitter only adds to the positioning error after the final stage.

The first stage of an ACQ/PEAK sequence generally uses the 4.3 aperture, which is larger than the height of the diode array. Hence, not just the positioning of the aperture on the sky, but also the positioning of the diode array on the photocathode determines which region of the source light is detected from. There are uncertainties of $\sim 0.1''$ in our knowledge of the correct $y$-locations of images and spectra on the photocathode. The geo-magnetic image motion problem (GIMP), which is not corrected for in ACQ/PEAK observations, can add to this further. However, simulations for the sequence in Table I indicate that this does not lead to acquisition failure. The target might not be contained in the aperture after the first stage, but the later stages always recover the target adequately, because of the overlap between subsequent stages.

## 4. Conclusions

The ACQ/PEAK acquisition mode can be used for a large variety of targets. However, its success is difficult to predict a priori, even for point sources. Software is presented that simulates ACQ/PEAK acquisitions in a Monte-Carlo manner. Results confirm that the standard exposure times should yield reliable acquisitions for point source targets. This need not be true for more complicated targets. Observers are advised to use the software to simulate their proposed acquisitions. It allows a proper quantitative assessment to be made of whether an acquisition will succeed, what exposure times should be used, and how the efficiency can be optimized.

## Acknowledgements

The author is grateful to the FOS Instrument Scientists Tony Keyes and Anuradha Koratkar for many helpful discussions. The author was supported by NASA through a Hubble Fellowship, #HF-1065.01-94A, awarded by the Space Telescope Science Institute which is operated by AURA, Inc., for NASA under contract NAS5-26555.

## Appendix

## A  Convolution of the target brightness

This appendix describes how the software integrates the PSF convolved target brightness over the size of the aperture.

### A1  THE CONVOLUTION KERNEL

Non-circularly symmetric features in the HST/FOS PSF are not of major importance in the present context. These tend to average out when integrating over the aperture size. We therefore restrict ourselves to circularly symmetric PSFs. It is assumed that the PSF can be expanded as a sum of Gaussians:

$$\text{PSF}(r) = \sum_{i=1}^{N} \frac{\gamma_i}{2\pi\sigma_i^2} \exp[-\frac{1}{2}(\frac{r}{\sigma_i})^2]. \tag{2}$$

The $\gamma_i$ and $\sigma_i$ are free parameters that can be chosen to optimize the fit to the actual PSF (see Appendix A.2 below). The $\gamma_i$ must satisfy $\sum_{i=1}^{N} \gamma_i = 1$, so that the PSF is normalized. The encircled flux $E(r)$ within a radius $r$ is given by

$$E(r) \equiv \int_0^r \text{PSF}(r')\, 2\pi r'\, dr' = 1 - \sum_{i=1}^{N} \gamma_i \, \exp[-\frac{1}{2}(\frac{r}{\sigma_i})^2]. \tag{3}$$

Hence, $1 - E(r)$ is also a sum of Gaussians.

Let $S(x, y)$ be the surface brightness of the source, with the coordinate system chosen along the $x$- and $y$-axes of the FOS. The PSF convolved surface brightness is

$$S_{\rm P}(x, y) = \int_{-\infty}^{\infty} \int_{-\infty}^{\infty} S(x', y') \, \text{PSF}(x' - x, y' - y)\, dx'\, dy'. \tag{4}$$

The most commonly used FOS apertures are either rectangular or circular. The total intensity, $I(x, y)$, observed through a rectangular aperture with sides $A_x$ and $A_y$, centered on $(x, y)$, is

$$I_{\rm rec}(x, y) = \int_{-A_x/2}^{A_x/2} \int_{-A_y/2}^{A_y/2} S_{\rm P}(x + x'', y + y'')\, dx''\, dy''. \tag{5}$$

The total intensity, $I(x, y)$, observed through a circular aperture with diameter $D$, centered on $(x, y)$, is

$$I_{\rm circ}(x, y) = \int_0^{2\pi} \int_0^{D/2} S_{\rm P}(x + r\cos\phi, y + r\sin\phi)\, r\, dr\, d\phi. \tag{6}$$

Substitution of $S_{\rm P}$ from eq. (4) and exchanging the order of the integrations yields:

$$I(x, y) = \int_{-\infty}^{\infty} \int_{-\infty}^{\infty} S(x', y')\, K(x' - x, y' - y)\, dx'\, dy'. \tag{7}$$

The convolution kernel $K(x, y)$ is for the rectangular aperture

$$K_{\rm rec}(x, y) = \int_{-A_x/2}^{A_x/2} \int_{-A_y/2}^{A_y/2} \text{PSF}(x - x'', y - y'')\, dx''\, dy'', \tag{8}$$

while for the circular aperture it is

$$K_{\rm circ}(x, y) = \int_0^{D/2} \int_0^{2\pi} \text{PSF}(x - r\cos\phi, y - r\sin\phi)\, r\, dr\, d\phi. \tag{9}$$

The transmission for a point source centered in the aperture is $K(0, 0)$, the convolution kernel evaluated at the origin.

With the PSF as in eq. (2), the double integral in eq. (8) reduces to

$$K_{\rm rec}(x, y) = \sum_{i=1}^{N} \frac{\gamma_i}{4} \left[ \text{erf}(\frac{x + (A_x/2)}{\sqrt{2}\,\sigma_i}) - \text{erf}(\frac{x - (A_x/2)}{\sqrt{2}\,\sigma_i}) \right]$$
$$\times \left[ \text{erf}(\frac{y + (A_y/2)}{\sqrt{2}\,\sigma_i}) - \text{erf}(\frac{y - (A_y/2)}{\sqrt{2}\,\sigma_i}) \right], \tag{10}$$

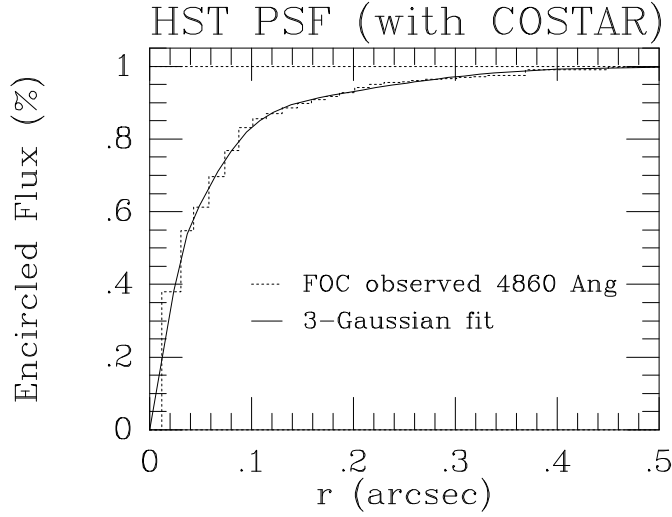

Fig. 3. The dotted curve shows the encircled flux for the HST PSF, as measured with the FOC (after the installation of COSTAR) at 4860Å. The solid curve is the best fit of a PSF which is the sum of three Gaussians, as described in the text.

where

$$\mathrm{erf}(x) \equiv \frac{2}{\sqrt{\pi}} \int_0^x e^{-t^2}\, dt. \qquad (11)$$

To evaluate the integral in eq. (9) define $x = \rho \cos \theta$ and $y = \rho \sin \theta$. Then

$$K_{\mathrm{circ}}(\rho \cos \theta, \rho \sin \theta) =$$

$$\sum_{i=1}^{N} \frac{\gamma_i}{2\pi \sigma_i^2} \exp[-\frac{1}{2}(\frac{\rho}{\sigma_i})^2] \int_0^{D/2} \exp[-\frac{1}{2}(\frac{r}{\sigma_i})^2]\, r\, dr \int_0^{2\pi} \exp[\frac{r\rho}{\sigma_i^2} \cos(\phi - \theta)]\, d\phi. \qquad (12)$$

The inner integral can be solved in terms of a Bessel function $I_0$, using eq. (3.915.4) of Gradshteyn & Ryzhik (1994). This yields

$$K_{\mathrm{circ}}(\rho \cos \theta, \rho \sin \theta) = \sum_{i=1}^{N} \frac{\gamma_i}{\sigma_i^2} \exp[-\frac{1}{2}(\frac{\rho}{\sigma_i})^2] \int_0^{D/2} \exp[-\frac{1}{2}(\frac{r}{\sigma_i})^2]\, I_0(\frac{r\rho}{\sigma_i^2})\, r\, dr. \qquad (13)$$

The remaining integral must be evaluated numerically.

## A2 Approximating the HST/FOS PSF

Few observations or models are available for the PSF of the FOS with COSTAR. The current implementation of the software is therefore based on an encircled flux curve measured with the FOC through a filter centered at 4860Å (FOC Instrument Handbook, Nota et al. 1994). This curve is displayed in Fig. 3 (dotted histogram). The solid curve is the best fit of eq. (3) with $N = 3$ Gaussians. The normalizations and dispersions of the Gaussians in this fit are: $\gamma_1 = 0.45282$, $\gamma_2 = 0.40878$, $\gamma_3 = 0.13840$, $\sigma_1 = 0.015''$, $\sigma_2 = 0.049''$, $\sigma_3 = 0.170''$.

ROELAND P. VAN DER MARELTABLE II

Aperture sizes and predicted transmissions for centered point sources

| aperture | x-size (arcsec) | y-size (arcsec) | diameter (arcsec) | — Absolute Transmission — | |
|---|---|---|---|---|---|
| | | | | our approx. PSF 4860Å | TIM PSF 5000Å |
| 4.3 | 3.698 | 3.698 | — | 1.000 | 0.98 |
| 0.25×2.0 | 0.215 | 1.72 | — | 0.915 | 0.87 |
| 1.0-PAIR | 0.86 | 0.86 | — | 0.997 | 0.95 |
| 0.5-PAIR | 0.43 | 0.43 | — | 0.949 | 0.90 |
| 0.25-PAIR | 0.215 | 0.215 | — | 0.869 | 0.79 |
| 0.1-PAIR | 0.086 | 0.086 | — | 0.611 | 0.57 |
| 1.0 | — | — | 0.86 | 0.994 | 0.94 |
| 0.5 | — | — | 0.43 | 0.938 | 0.89 |
| 0.3 | — | — | 0.258 | 0.883 | 0.84 |

The PSF for the FOS need not be the same as for the FOC. FOS aperture throughputs for centered point sources were calculated recently at STScI using the TIM software (Table 1–8 of the FOS Instrument Handbook). The predictions at 5000Å can be compared to the predictions obtained with our fit to the PSF (valid at $\sim 4860$Å). The results are listed in Table II. The aperture sizes based on a COSTAR reduction factor of 0.86 are also listed. The transmissions predicted by our model are on average 5% larger than those calculated with the TIM software, with little dependence on the size of the aperture. This discrepancy is due to differences in the PSF at large radii. In our PSF model hardly any light is observed at radii $\gtrsim 0.5''$. In the TIM PSF as much as 5% of the light is scattered to radii $\gtrsim 0.5''$. Apart from this, our model captures the essence of the PSF convolution well.

The PSF depends in a non-trivial way on wavelength. The current implementation of the software does not correct for this. Instead, a 'fudge factor' $F$ is provided which changes the FWHM of the multi-Gaussian PSF while leaving its shape unchanged (i.e., $\sigma_i \to F \times \sigma_i$). The user can choose this factor so as to optimize the agreement between the predicted aperture transmissions for a centered point source and the known aperture transmissions at the wavelength of interest (consult the FOS Instrument Handbook). The factor $F$ can also be used to assess the extent to which the success of the ACQ/PEAK sequence depends on the details of the PSF.

The current implementation of the PSF in the software suffices for most situations. However, the parameters of the multi-Gaussian PSF (including the number of Gaussians) can easily be updated in the software when better approximations to the PSF become available.

*Calibrating HST: Post Servicing Mission*